\documentclass[12pt]{article}

\pdfoutput=1

\usepackage{amsmath,amssymb,amsfonts,amscd,mathrsfs}
\usepackage{xcolor}
\definecolor{darkblue}{rgb}{0.1,0.1,.7}
\usepackage[colorlinks, linkcolor=darkblue, citecolor=darkblue, urlcolor=darkblue, linktocpage]{hyperref} 
\usepackage[square, comma, compress,numbers]{natbib}
\usepackage[]{graphicx}
\usepackage{geometry}
\usepackage[margin=10pt,font=small,labelfont=bf]{caption}
\usepackage{ifthen}
\usepackage{tikz}
\usepackage{subcaption}
    
\usepackage{dsfont} 

\usepackage{titlesec}
\titleformat*{\section}{\large\bfseries}
\titleformat*{\subsection}{\normalsize\bfseries}
\titleformat*{\subsubsection}{\normalsize\bfseries}
\titleformat*{\paragraph}{\normalsize\bfseries}
\titleformat*{\subparagraph}{\normalsize\bfseries}

\def\Emaxbar{\bar{E}_{\rm max}}

\newcommand{\reef}[1]{(\ref{#1})}

\newcommand{\beq}{\begin{equation}} 
\newcommand{\eeq}{\end{equation}}
\def\del {\partial} 
\def\nn{\nonumber} 
\def\bZ {\mathbb{Z}}

\def\calN {{\cal N}} 
 
\def\calH {{\cal H}}

\def\calA {{\cal A}} 
 
\def\calE {{\cal E}} 
\def\bZ {\mathbb{Z}} 
 
\def\bP {\mathbb{P}} 
\def\half{{\textstyle\frac 12}}

\def\ge{\geqslant}
\def\le{\leqslant}

\newcommand{\diffop}[2]{\ifthenelse{\equal{#2}{1}}{\frac{\mrm{d}}{\mrm{d} #1}}{\frac{\mrm{d}^#2}{\mrm{d} #1^#2}}}
\newcommand{\NO}[1]{{:\!#1\!:}}

\newcommand{\braket}[3]{\langle #1|#2|#3 \rangle}

\newcommand{\ket}[1]{|#1\rangle}
\newcommand{\bra}[1]{\langle #1|}

\newcommand{\mrm}[1]{{\mathrm #1}}

\usepackage[normalem]{ulem}

\def\del{\partial}

\def \Emax{E_{\rm max}}
\def \Htr{{H_{\rm trunc}}}

\usepackage{accents}
\newlength{\dhatheight}

\numberwithin{equation}{section}
\begin{document}

\vspace*{-.6in} \thispagestyle{empty}
\begin{flushright}
CERN PH-TH/2015-277\\
\end{flushright}
\vspace{1cm} {\Large
\begin{center}
{\bf Hamiltonian Truncation Study of the $\phi^4$ Theory\\ 
in Two Dimensions\\
II. The $\bZ_2$-Broken Phase and the Chang Duality}\\
\end{center}}
\vspace{1cm}
\begin{center}
{\bf Slava Rychkov$^{a,b,c}$,  Lorenzo G. Vitale$^d$}\\[2cm] 
{
$^{a}$ CERN, Theory Department, Geneva, Switzerland\\
$^{b}$ Laboratoire de Physique Th\'{e}orique de l'\'{E}cole normale sup\'{e}rieure (LPTENS), Paris, France\\
$^{c}$ Sorbonne Universit\'es, UPMC Univ Paris 06, Facult\'e de Physique, Paris, France\\
$^d$ Institut de Th\'eorie des Ph\'enom\`enes Physiques, EPFL, CH-1015 Lausanne, Switzerland\\
}
\vspace{1cm}
\end{center}

\vspace{4mm}

\begin{abstract}
The Fock-space Hamiltonian truncation method is developed further, paying particular attention to the treatment of the scalar field zero mode.
This is applied to the two-dimensional $\phi^4$ theory in the phase where the $\bZ_2$-symmetry is spontaneously broken, complementing our earlier study of the $\bZ_2$-invariant phase and of the critical point. We also check numerically the weak/strong duality of this theory discussed long ago by Chang.
 \end{abstract}
\vspace{.2in}
\vspace{.3in}
\hspace{0.7cm} December 2015

\newpage

{
\tableofcontents
}
\newpage

\setlength{\parskip}{0.1in}

\section{Introduction}

The two-dimensional $\phi^4$ theory is perhaps the simplest quantum field theory (QFT) which is not exactly solvable. It is thus an ideal laboratory for studying approximate solution techniques. In our recent paper \cite{Rychkov:2014eea}, we studied this theory using the method of \emph{Hamiltonian truncation}---a QFT analogue of the Rayleigh-Ritz method in quantum mechanics. In that work, we considered the case of positive bare mass $m^2>0$ and of quartic coupling $g=\bar g m^2$ with $\bar g=O(1)$.
The physical particle mass is given by
\beq
m_{\rm ph} = f(\bar g) m\,,
\eeq
and the function $f(\bar g)$ was determined numerically. We observed that the physical mass vanishes for $\bar g=\bar g_c\approx 3$,
signaling the presence of a second order phase transition. 

In \cite{Rychkov:2014eea}, our focus was mainly on the region below and around the critical coupling $\bar g_c$.  In this second work of the series we will instead be interested in the complementary region $\bar g>\bar g_c$. In this {range of couplings} the theory is massive, but the $\bZ_2$ symmetry, $\phi\to-\phi$, is spontaneously broken. In infinite volume, there are therefore two degenerate vacua, and two towers of massive excitations around them. 

We will be able to determine the low energy spectrum as a function of $\bar g$. In finite volume the exact degeneracy is lifted, and the energy eigenstates come in pairs split by a small amount, exponentially small if the volume is large. In this paper, as in \cite{Rychkov:2014eea}, we will regulate the theory by putting it in finite volume.

In the $\bZ_2$-broken phase, there is also a topologically nontrivial sector of ``kink" states corresponding, in the semiclassical limit, to field configurations interpolating between the two vacua. In this work we will probe the kink mass by studying the mass splittings in the topologically trivial sector. In the future it would be interesting to study the kink sector directly.

One interesting feature of the theory under study is that it enjoys a weak/strong coupling duality first discussed
by Chang \cite{Chang:1976ek}. The dual description exists for all $\bar{g}\ge \bar{g}_*\approx 2.26$. As we review below, the duality relates a description in which the theory is quantized around the $\bZ_2$-invariant vacuum state to an equivalent description in which it is quantized around a $\bZ_2$-breaking vacuum. For $\bar{g}$ not much above $\bar{g}_*$ both descriptions are strongly coupled\footnote{This explains why $\bar{g}_*$ need not be equal, and in fact is not equal to the critical coupling $\bar{g}_c$ mentioned above.} and they can be equivalently
employed as a starting point for the numerical computations. In section \ref{sec:chang}  we present a comparison between the numerical spectra obtained using the two descriptions, serving both as a non-trivial test of the method and as a check of the Chang duality.

On the other hand, for $\bar{g} \gg \bar{g}_*$ the dual description becomes weakly coupled, and provides the better starting point.  
In section \ref{sec:weakcoupling}, we will explain a modification of the method which can be used, among other things, to study this regime (a weakly coupled $\phi^4$ theory with negative $m^2$) efficiently. It is based on a different treatment of the zero mode of the field. We will compare the numerical results with the predictions from perturbation theory and from semiclassical analyses.

We conclude in section \ref{sec:conclusions}. Several technical details are relegated to the appendices.

Recently, the $\bZ_2$-broken phase of the two-dimensional $\phi^4$ model was studied in Ref.~\cite{Coser:2014lla} using a version of the Truncated Conformal Space Approach \cite{Yurov:1989yu,Lassig:1990xy}. Differences and similarities between our works will be mentioned throughout the paper. 

\section{The Chang duality}
\label{sec:chang}

\subsection{Formulation and consequences}

According to Chang \cite{Chang:1976ek}, the two-dimensional $\phi^4$ theory described by the (Euclidean) Lagrangian
\beq
\mathcal{L}=
\half (\del\phi)^2 +\half m^2 \phi^2 + g\, N_m(\phi^4)
\label{eq:L}
\eeq
with $m^2>0$, $g>0$, admits a dual description in terms of a Lagrangian with a different, and negative, value of the squared mass:
\beq
\mathcal{L'}=
 \half (\del\phi)^2 - {{\textstyle\frac 14}}
M^2 \phi^2 + g\, N_M(\phi^4)\,.
 \label{eq:L'}
\eeq
The actual value of the dual mass will be given below.

Note that the duality is between quantum theories in the continuum limit, and to specify this limit one has to subtract the logarithmic divergence of the mass parameters. The divergence is removed by normal-ordering the quartic interaction with respect to the mass indicated in the subscript of the normal ordering sign $N$. The potential in $\mathcal{L'}$ has two minima at $\phi=c=\pm M/\sqrt{8g}$. After the shift $\phi\to\phi+c$ the dual Lagrangian becomes\footnote{Notice that normal ordering is a linear operation, and thus commutes with the field shift.} 
\beq
\label{eq:shiftedpotential}
\mathcal{L'} \to  \half (\del\phi)^2 + \half M^2 \phi^2 + \sqrt{2g}M\, N_M(\phi^3)  + g\, N_M(\phi^4)\,.
\eeq
In this way of writing, interactions of both $\mathcal{L}$ and $\mathcal{L}'$ are normal ordered with respect to the mass appearing in the quadratic part of the Lagrangian. In perturbation theory such normal ordering means that we are simply forbidding diagrams with the lines starting and ending in the same vertex.

\begin{figure}[h]
\begin{center}
  \includegraphics[scale=0.4]{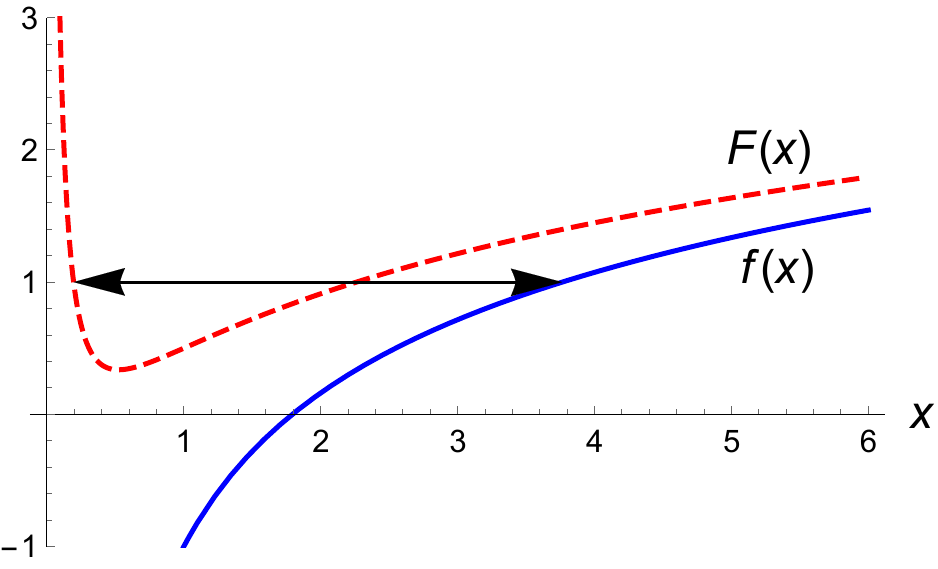}\hspace{1cm}
  \includegraphics[scale=0.4]{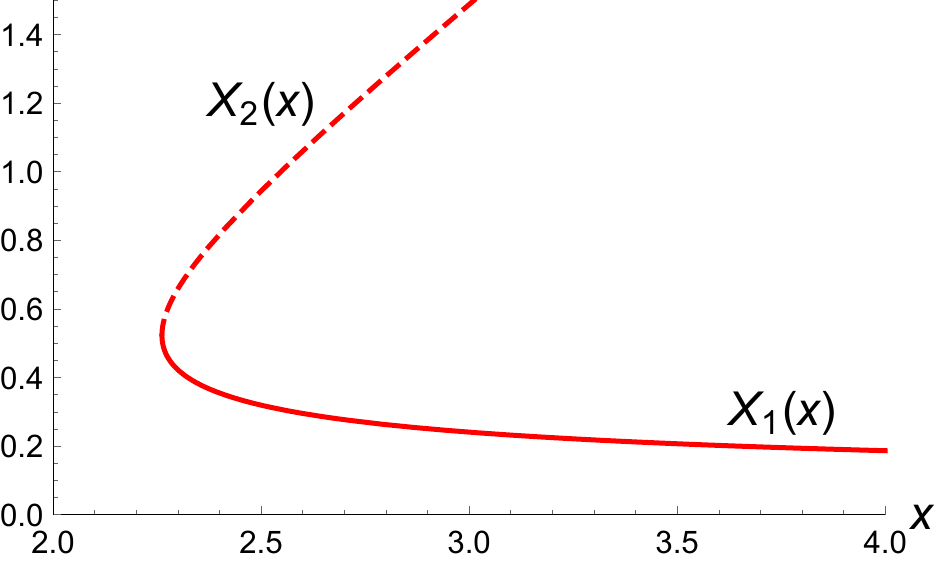} 
  \end{center}
\caption{Left panel: equation $F(X)=f(x)$ has two solutions for $x>x_*$. Right panel: the two solution branches $X_{1,2}(x)$. We are mostly interested in the lower branch $X_1(x)$ which becomes weakly coupled as $x\to\infty$.  }
\label{fig:eq}
\end{figure}

To find the dual mass $M^2$, one is instructed to solve the equation: 
\begin{gather}
\label{eq:ch0}
F(X)=f(x)\,,
\end{gather}
where $x=g/m^2$, $X=g/M^2$ are the dimensionless quartic couplings of the two descriptions ($x$ is given and $X$ is an unknown) and
\beq
f(x)\equiv\log x-\pi/(3 x)\,,\qquad F(X)\equiv\log X+\pi/(6X)\,.
\eeq 
This equation is illustrated in Fig.~\ref{fig:eq}. There is no solution for 
\beq
x<x_*=\frac{\pi}{3 W(2/e)}\approx 2.26149\,,
\eeq 
where $W(z)$ is the Lambert $W$ function. For $x\ge x_*$ there are two solution branches.
We are particularly interested in the lower branch $X_1(x)$, which for large $x$ approaches zero:
\beq
X_1(x)\approx \pi/(6 \log x),\quad x\to \infty\,.
\eeq 
The dual description corresponding to this branch becomes weakly coupled in the limit in which the original description becomes stronger and stronger coupled. We thus have a weak/strong coupling duality.

Chang \cite{Chang:1976ek} used this duality to show that the $\phi^4$ theory undergoes a phase transition. Indeed, for small $x$ we can use perturbation theory to argue that the theory is in the symmetric phase, with the $\bZ_2$ symmetry $\phi\to -\phi$ unbroken. On the other hand, for large $x$ we use the dual description. Since in that description the potential is a double well, and moreover the dual coupling is weak for $x\gg 1$, we conclude that for large $x$ the $\bZ_2$ symmetry is spontaneously broken. By continuity, there must be a phase transition at an intermediate value of $x$. 

This argument does not establish whether the transition is first or second order. However, as explained in \cite{Chang:1976ek}, a first order transition is excluded by rigorous 
theorems due to Simon and Griffiths \cite{Simon:1974dg}. So the transition must be second order. This conclusion is supported by Monte Carlo simulations \cite{Loinaz:1997az,Schaich:2009jk,Wozar:2011gu,Bosetti:2015lsa}, as well as by computations using DLCQ \cite{Harindranath:1987db}, density matrix renormalization group \cite{Sugihara:2004qr}, matrix product states \cite{Milsted:2013rxa}, and the Hamiltonian truncation \cite{Lee:2000ac,Lee:2000xna,Rychkov:2014eea}.

Nor does the above argument predict the value of $x$ at which the phase transition must happen. In particular, the fact that the dual description exists at $x\ge x_*$ does not mean that the phase transition happens at $x=x_*$. Indeed, at $x=x_*$ both the direct and the dual descriptions are strongly coupled, and the fate of the $\bZ_2$ symmetry is not \emph{a priori} clear. In fact, calculations indicate a higher phase transition location at $x_c\approx 2.75 - 3$ \cite{Rychkov:2014eea,Wozar:2011gu,Milsted:2013rxa,Bosetti:2015lsa,Pelissetto:2015yha}.

\subsection{Review of the derivation}
Here's a quick derivation of the Chang duality, following \cite{Chang:1976ek}. We will work in the Hamiltonian formalism, and consider the normal-ordered Hamiltonians corresponding to $\mathcal{L}$ and $\mathcal{L'}$:
\begin{gather}
H=
\int dx\, N_m\bigl(\half \dot \phi^2 + \half \phi'^2 + \half m^2 \phi^2 +g\, \phi^4\bigr)
\label{eq:H}\,,\\
H'=
\int dx\, N_M\bigl(\half \dot \phi^2 + \half \phi'^2 -{{\textstyle\frac 14}} M^2 \phi^2 + g\,\phi^4+\Lambda\bigr)\,.
\label{eq:H'0}
\end{gather}
Notice that we are now normal ordering the full Hamiltonian, including the quadratic part. This more careful procedure will allow us to establish the correspondence also for the ground state energy. In the dual description it will receive an extra constant contribution, denoted~$\Lambda$ in \reef{eq:H'0}.

Recall Coleman \cite{Coleman:1974bu} relations between normal orderings with respect to different masses:
\begin{gather}
N_m\bigl(\half \dot \phi^2 + \half \phi'^2) = N_M\bigl(\half \dot \phi^2 + \half \phi'^2)+ Y\,,\nn\\
N_m(\phi^2) = N_M(\phi^2) + Z\,,\label{eq:coleman}\\
N_m(\phi^4) = N_M(\phi^4) + 6 Z N_M(\phi^2)+ 3 Z^2\,,
\nn
\end{gather}
where $Y = Y(m,M)$ and $Z = Z(m,M)$ are the differences of the normal-ordering constants:\footnote{The expression for $Z$ can also be equivalently derived in the Lagrangian language as the difference of one-loop massive diagrams: 
$
Z = \int \frac{d^2 k}{(2\pi)^2} \bigl( \frac{1}{k^2 + M^2} - \frac{1}{k^2 + m^2} \bigr)\,.
$}
\begin{gather}
\label{eq:YZ}
Y(m,M)=\int \frac{dk}{8\pi}\left\{
\frac{2 k^2+M^2}{\sqrt{k^2+M^2}}-(M\to m)\right\}=\frac 1{8\pi}(M^2 - m^2)\,,\nn\\
Z(m,M)=\int \frac{dk}{4\pi}
\left\{
\frac{1}{\sqrt{k^2+M^2}}-(M\to m)\right\}=\frac 1{4\pi}\log \frac{m^2}{M^2}\,.
\label{eq:YZ}
\end{gather}
Using these relations, one can see that $H$ maps on $H'$ as long as 
\beq
\label{eq:M2}
\half m^2+6 Z g =-{{\textstyle\frac 14}} M^2\,,
\eeq
written equivalently as \reef{eq:ch0}. We also find a constant contribution to the ground state energy
\beq
\label{eq:Lambda}
\Lambda = Y +\half m^2 Z +{3 g} Z^2\,.
\eeq

\subsection{Numerical check of the duality}
\label{sec:setup}

We will test the Chang duality by comparing the spectra of the direct and dual theories in a finite volume---a circle of length $L$. The spectra will be computed using the Hamiltonian truncation. We will first describe the setup for these computations, and then present the results.

\subsubsection{Direct theory}
\label{sec:directtheory}

By the direct theory we mean \reef{eq:H} put on a circle of length $L$. This is precisely the theory we were studying in \cite{Rychkov:2014eea}, and we will be following the same method. Here we will give just a brief reminder. 
The finite volume Hamiltonian corresponding to the infinite-volume Hamiltonian \reef{eq:H} is given in \cite{Rychkov:2014eea}, Eq.~(2.19), and has the form:
\beq
H(L) = H_0 + g[V_4+ 6 \zeta V_2]+ [E_0+3 \zeta^2 g L],
\label{eq:HL}
\eeq
Here $H_0$ is the Hamiltonian of the free scalar field on the circle:
\beq
H_0(L,m)=\sum_k \omega(k) a^\dagger_k  a_k,\quad k=(2\pi/L)n,\ n\in\bZ,\quad \omega(k)=\sqrt{m^2+k^2}\,,
\eeq
where $a,a^\dagger$ are the ladder operators appearing in the field mode expansion:\
\begin{equation}
\label{eq:modeexp}
\phi(x) = \sum_{k} \frac{1}{\sqrt{2L  \omega(k)}} \left( a_k e^{i k x} + a_k^\dagger e^{-i k x}\right)\,.
\end{equation}
The $V_4$ term is the normal ordered quartic interaction:
\beq
V_4(L,m) = \frac{1}{L} \sum_{\sum k_i=0} \frac{1}{\prod \sqrt{2 \omega(k_i)}}
\Big[ (a_{k_1}a_{k_2}a_{k_3}a_{k_4} +
4 a^\dagger_{-k_1}a_{k_2}a_{k_3}a_{k_4} +\text{h.c.}) +
6 a^\dagger_{-k_1}a^\dagger_{-k_2}a_{k_3}a_{k_4} 
 \Big]\,.
\eeq
The other terms in \reef{eq:HL} are all exponentially suppressed for $Lm\gg 1$. In particular,
\beq
\label{eq:E0L}
E_0(L,m)=-\frac{1}{\pi L}\int_0^\infty dx \frac{x^2}{\sqrt{m^2L^2+x^2}} \frac{1}{e^{\sqrt{m^2L^2+x^2}}-1}\, .
\eeq
is the Casimir energy of the free scalar field in finite volume. Corrections involving
\beq
\zeta(L,m)=\frac{1}{\pi }\int_0^\infty \frac{dx}{\sqrt{m^2L^2+x^2}}\frac{1}{e^{\sqrt{m^2L^2+x^2}}-1}\, .
\eeq
are due to a mismatch between the normal ordering counterterms needed to define the $\phi^4$ operator in infinite space and on the circle. One of them contributes to the vacuum energy density, and the other is a correction proportional to the mass operator $V_2$: 
\beq
V_2(L,m)=\sum_k \frac{1}{2 \omega_k}( a_k a_{-k} + a^\dagger_k a^\dagger_{-k} +2 a^\dagger_k a_k)\,.
\eeq 
In \cite{Rychkov:2014eea} we worked at circle sizes up to $L=10 m^{-1}$, and it was justified to neglect the exponentially small terms proportional to $E_0$ and $z$. 
Here, in some cases, we will work at smaller circle sizes. In this paper we will 
always keep these terms, which is actually straightforward in our algorithm.

The Hilbert space $\calH$ of the theory is the Fock space of the ladder eigenstates. As in \cite{Rychkov:2014eea}, we will restrict our attention to the subsector of the Hilbert space consisting of the states of zero total momentum $P=0$ and of the positive spatial parity $\bP=1$. The Hamiltonian \reef{eq:HL} does not mix states of positive and negative field parity $\bZ_2:\phi\to-\phi$ (i.e.~the states containing an even and odd number of particles). Thus the $\bZ_2$-even and $\bZ_2$-odd sectors can be studied separately. We will study both of them. Finally, we will truncate the Hilbert space to the subspace of states $\calH(\Emax)$ which have $H_0$ energy below a certain cutoff $\Emax$. Typically, we choose our cutoff so that the dimension of $\calH(\Emax)$ is $\sim$10000 per $\bZ_2$ sector. The Hamiltonian $H(L)$ restricted to the truncated Hilbert space is called the truncated Hamiltonian $H(L)_{\rm trunc}$.

We evaluate the matrix elements of $\Htr$, and the eigenvalues of the resulting finite matrix are then computed numerically. This gives what in \cite{Rychkov:2014eea} is called ``raw" spectrum. It converges to the true nonperturbative spectrum with a rate which asymptotically goes as $1/\Emax^2$. 

Convergence of the method can be improved by renormalizing the couplings. We refer the reader to \cite{Rychkov:2014eea} for a detailed explanation of the renormalization procedure.\footnote{Similar renormalization procedures were developed in the Truncated Conformal Space Approach literature \cite{Giokas:2011ix,Feverati:2006ni,Watts:2011cr,Lencses:2014tba}. The concrete version used by us shares a lot in common with the one in \cite{Hogervorst:2014rta}; the small differences that exist were stressed in \cite{Rychkov:2014eea}.} In the present work we will use an identical procedure, apart from a technicality that we now explain.

In \cite{Rychkov:2014eea}, the leading renormalization coefficients were calculated by extracting the leading non-analytic behavior for $\tau \to 0$ of the quantities
\begin{equation}
\label{eq:ik}
I_k(\tau) = \int_{- L/2}^{L/2} d z\, G_L(z,\tau)^k\,,
\end{equation}
where $G_L(z,\tau)$ is the two point function in finite volume, which can be expressed through periodization via the two point function in infinite volume:
\begin{gather}
G_L (z, \tau) = \sum_{n \in \mathds{Z}} G(\sqrt{(z+n L)^2 + \tau^2})\,, \label{eq:gl}
\\ G(\rho) \equiv \frac{1}{2 \pi} K_0(m \rho) \,, \quad \rho \equiv \sqrt{z^2+\tau^2}\,.
\end{gather}
Here $K_0(m \rho)$ is a modified Bessel function of the second kind. Since $G(\rho)$ is exponentially suppressed for $m \rho \gg 1$, the contributions from $n\ne 0$ in (\ref{eq:gl}) can be neglected as long as $m L \gg 1$. This is what we did in \cite{Rychkov:2014eea}. However in the present work we will encounter also the situation $m L = O(1)$. Our procedure will be to approximate:
\begin{equation}
\label{eq:approxgl}
G_L(z, \tau) \simeq  G(\rho) + 2 \sum_{n=1}^{\infty} G(n L)\,,
\end{equation}
which simply adds a constant to the infinite-volume two point function. This approximation is justified because the higher order Taylor expansion terms of $G(\rho)$ around $\rho = n L$ would result in renormalization terms suppressed by powers of $m^2/\Emax^2 \ll 1$. 
The short-distance asymptotics of $G_L$ used to calculate (\ref{eq:ik}) is modified as (cf.~(3.23) in \cite{Rychkov:2014eea}):
\begin{equation}
G_L(z,\tau) \approx - \frac{1}{2 \pi} \log \left(\frac{e^\gamma}{2} m' \rho \right) \,, \quad m' \equiv  m \exp\Bigl[- 4 \pi \sum^\infty_{n=1} G(n L)\Bigr]\,.
\end{equation}

It is then straightforward to generalize the renormalization procedure of \cite{Rychkov:2014eea} to the case $m L =O(1)$.
E.g.~the Hamiltonian renormalized by local counterterms is given by:
\beq
\label{eq:Hren}
H(L)_{\rm ren}=H_{\rm trunc}(L)+\int dx\, N_m(\kappa_0 +\kappa_2 \phi^2 + \kappa_4 \phi^4)\,,
\eeq
where $\kappa_i$ are given in \cite{Rychkov:2014eea}, (3.34) where one has to put $g_4=g$, $g_2=6z(L)g$, and replace $m\to m'$ in the expressions for the $\mu$-functions in \cite{Rychkov:2014eea}, (3.31).
This Hamiltonian allows to calculate the spectrum with the convergence rate of $1/\Emax^3$. In the numerical computations in section \ref{sec:result} we will also include subleading, non-local corrections improving the convergence rate up to $1/\Emax^4$, for which we refer the reader to \cite{Rychkov:2014eea}.

\subsubsection{Dual theory}
\label{sec:dualtheory}
 
The Hamiltonian for the dual theory in finite volume is easiest derived as follows. 
Let us rewrite $H'$ in \reef{eq:H'0} by adding and subtracting $\half M^2\phi^2$:
\beq
\label{eq:H'}
H'=
\int dx\, N_M\bigl(\half \dot \phi^2 + \half \phi'^2 + \half M^2\phi^2\bigr) +N_M\bigl(-{{\textstyle\frac 34}} M^2 \phi^2 + g\,\phi^4+\Lambda\bigr)\,.
\eeq
This looks like the direct Hamiltonian with $m\to M$ and an extra negative mass squared perturbation. The passage to a finite volume is then analogous to the direct theory. We get
\begin{gather}
H'(L) = H_0 + \Bigl [ -{{\textstyle\frac 34}} M^2 + 6 \zeta g\Bigr] V_2+ g V_4+ h\,, \label{eq:H'L0} \\
h =  \Lambda L+ E_0+3 \zeta^2 g L 
-{{\textstyle\frac 34}} M^2 \zeta L.
\label{eq:H'L}
\end{gather}
The building blocks have the same meaning as in section \ref{sec:directtheory}, except that we have to use $M$ instead of $m$ in all expressions: $H_0=H_0(L,M)$, $\zeta=\zeta(L,M)$, etc.

\begin{figure}[h!]
\begin{center}
\includegraphics[scale=.8]{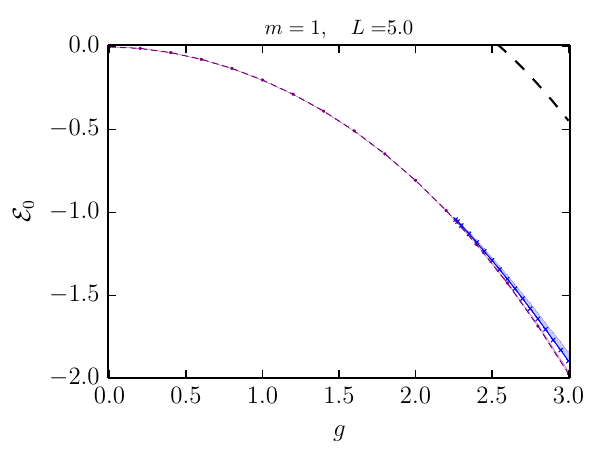}
\includegraphics[scale=.8]{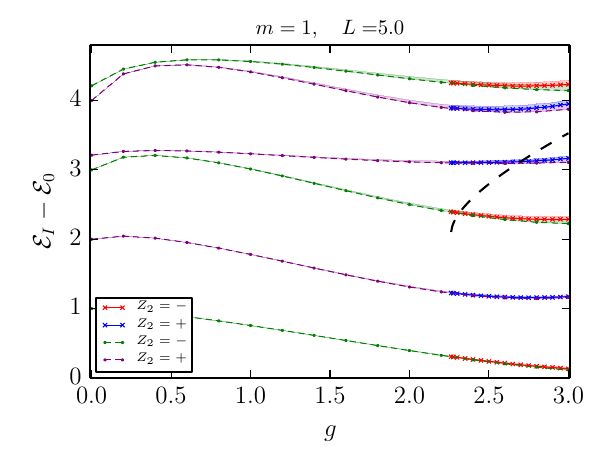} 
\end{center}
\caption{The ground state energy (left) and the spectrum of excitations (right) for the direct and the dual theory as a function of $g$ for $m=1$, $L=5$. The excitation plot shows the energies of the $\bZ_2$ odd and $\bZ_2$ even energy levels. See the text for the details.}
\label{fig:comparison}
\end{figure}

\subsubsection{Comparison}
\label{sec:result}

In figure \ref{fig:comparison} we show the ground state energy $\calE_0$ and the spectrum of excitations $\calE_I-\calE_0$ for $m=1$, $L=5$. We plot them as a function of the direct coupling $g=0$ - 3. The results for the direct theory are given in the full range of $g$, whereas for the dual theory {only} for $g\ge g_c\approx 2.26$, where the dual description exists. As in \cite{Rychkov:2014eea}, the error (shaded region) is estimated as the variation of the results upon using the ``local'' and ``subleading'' renormalization prescriptions. 

We see that in the overlapping region the {numerical predictions} from the two descriptions agree very well. This is an explicit check of the Chang duality. This check is non-trivial, as in both descriptions the Hamiltonian is strongly coupled. To illustrate this, the black dashed lines in the plots represent the tree-level prediction for the vacuum energy and the lightest excitation in the dual description. 

\emph{Computational details:} The computation in the direct theory is carried out as described in section \ref{sec:directtheory}. The dual mass $M$ for a given $g\ge g_c$ is determined by solving Eq.~\reef{eq:ch0} numerically. We use the solution with the smaller $X$ (and thus the larger $M$). The computation in the dual theory is then done using the Hamiltonian \reef{eq:H'L0} with two couplings $g_2=-{{\textstyle\frac 34}} M^2 + 6 z(L) g$ and $g_4=g$, i.e.~by including $-{{\textstyle\frac 34}} M^2$ into the perturbation. The renormalization procedure in \cite{Rychkov:2014eea} is applicable for such a general perturbation. It's not a problem for the method that $g_2$ is negative and comparable in size to the positive mass square term in $H_0$. There is in fact a great deal of arbitrariness in how to split the $\phi^2$ coefficient between the zeroth-order Hamiltonian and the perturbation. What we do here is just the fastest possibility, which turns out sufficient for the purposes of this section. More sophisticated ways of dealing with the dual theory will be developed in section \ref{sec:weakcoupling}.

\section{The $\bZ_2$-broken phase}
\label{sec:weakcoupling}

In section \ref{sec:chang} we reviewed the Chang duality and tested it numerically in the strongly coupled region by comparing the results obtained from the dual and {the} direct {descriptions}. We will now focus on the region $g/m^2 \gg g_*/m^2$, where the theory is in the $\bZ_2$-broken phase. In this {range of couplings} the direct description is very strongly coupled and it's difficult to achieve good numerical accuracy. On the other hand, the dual Hamiltonian becomes weakly coupled ($g/M^2\ll 1$). Therefore, we will use the dual Hamiltonian (\ref{eq:H'0}) as the starting point for the numerical calculations. 
It will be convenient to replace the value of $\Lambda$ given in (\ref{eq:Lambda}) by $\Lambda = {M^2}/({64 g})$, which corresponds to having zero classical vacuum energy density of the dual Hamiltonian.

\subsection{Modified zero mode treatment}
\label{sec:numres}

In \ref{sec:directtheory} we reviewed the method of \cite{Rychkov:2014eea} which treats all field modes on equal footing. This method is adequate in the $\bZ_2$-unbroken phase and in the $\bZ_2$-broken phase in moderate volumes, as in section \ref{sec:result}. However, it becomes inefficient in the $\bZ_2$-broken phase in large volume. The physical reason is that the zero mode has then very different dynamics from the rest of the modes, acquiring a VEV. It makes sense to take this into account, and to treat the zero mode separately from the rest. We will now explain how this can be done.

First of all we will rewrite \reef{eq:H'L0} making explicit the dependence on the zero mode. We will revert for the zero mode from using the oscillators $a_0,a_0^\dagger$ to the field variable
\beq
\phi_0=(a_0+a_0^\dagger)/\sqrt{2 L M}
\eeq
and the corresponding conjugate momentum $\pi_0$:
\begin{equation}
\pi_0 = i  (a_0^\dagger -a_0) \sqrt{L M/2}\,.
\end{equation}
Denoting by bar (resp.~hat) all quantities involving only the nonzero (zero) modes, we have
\begin{gather}
H_0 = \bar H_0 + \frac{\NO{\pi_0^2}}{2L}+ \frac{L M^2}{2} \NO{\phi_0^2 } \,,\\
V_2 = \bar V_2 + L \NO{\phi_0^2} \,, \quad V_4 = \bar V_4 + 4 \bar V_3 \phi_0 + 6 \bar V_2 \NO{\phi_0^2} + L \NO{\phi_0^4}\,.
\end{gather}
Gathering everything we get
\beq
\label{eq:ham0mode}
H'(L) =  \bar{H_0} + \hat{H} + W \,,
\eeq
where $\hat{H}$ depends only on the zero mode:
\beq
\label{eq:ham0mode2}
\hat{H} \equiv \frac{\NO{\pi_0^2}}{2L}+ L\left[ -{\textstyle \frac{1}{4}} M^2 + 6 \zeta g \right]\NO{\phi_0^2} + L g\, \NO{\phi_0^4} +h \,,
\eeq
while $W$ involves the interactions between the zero and the nonzero modes, and among the latter:
\beq
W \equiv \Bigl [ 6 g\NO{\phi_0^2}-{{\textstyle\frac 34}} M^2 + 6 \zeta g\Bigr] \bar V_2+ 4g \phi_0 \bar V_3+ g \bar V_4\,.
\label{eq:W}
\eeq

In a large volume and for $g \ll M^2$, the quantum mechanics of (\ref{eq:ham0mode2}) predicts that the wavefunction of $\phi_0$ is peaked around the minima of the potential at $\phi_0^2\approx M^2/(8g)$, with a width scaling asymptotically as $\langle (\Delta \phi_0)^2 \rangle \sim 1/(L M)$.
 For this $\phi_0$ the coefficient of ${\bar V}_2$ in $W$ vanishes. Intuitively this implies that, up to small perturbative corrections induced by the ${\bar V}_3$ and ${\bar V}_4$ terms, the nonzero modes of the field will stay in their vacuum state. This is true in a very large volume, and it provides a good starting point for a quantitative description in finite volume.

The idea of the method will be therefore to first solve the quantum mechanics of the zero modes, by neglecting its interaction with the nonzero modes. Having done so, the full Hamiltonian will be diagonalized in a Hilbert space whose basis wavefunctions are products of the exact zero mode wavefunctions and the harmonic oscillator wavefunctions for the nonzero modes. This is expected to be more efficient than the original method which would use harmonic oscillator wavefunctions also for the zero mode. 

Concretely, the procedure goes as follows. The full Hilbert space can be written as a direct product:
\begin{equation}
\mathcal{H} = \mathcal{\hat{H}} \otimes \mathcal{\bar{H}}\,,
\end{equation}
where $\mathcal{\hat{H}}$ and $\mathcal{\bar{H}}$ are the Hilbert spaces of the zero modes and nonzero modes, respectively. 
The truncated Hilbert space is then ($l$ for low)
\beq
\calH_l = \mathcal{\hat{H}}_l \otimes \mathcal{\bar{H}}_l\,,
\eeq
where the basis of $\mathcal{\bar{H}}_l$ is formed by the harmonic oscillator states for the nonzero modes with energy $\bar{E} \le \Emaxbar$, while $\mathcal{\hat{H}}_l$ is spanned by the first few low-lying eigenfunctions of $\hat{H}$:
\begin{equation}
\hat{H} \ket{\psi_\alpha} = \hat{\mathcal{E}}_\alpha \ket{\psi_\alpha},\quad \alpha=1\ldots s\,.
\end{equation}
In practice, it will be sufficient to fix $s=4$ or 5.

A separate computation has to be done to find the $\ket{\psi_\alpha}$. We do this using the standard Rayleigh-Ritz method, working in the $S$-dimensional subspace of $\mathcal{\hat{H}}$ spanned by the original harmonic oscillator wavefunctions $(a^\dagger_0)^i \ket{0}$, $i=0\ldots S-1$. The parameter $S\gg s$ can be chosen so large that the numerical error accumulated in this step is insignificant; in practice we choose $S=500$. The eigenstates $\ket{\psi_\alpha}$ are thus found expanding them in the harmonic oscillator wavefunctions. This facilitates the subsequent computations of the matrix elements involving these states.

One can now compute the matrix elements of $H'(L)$ in the truncated Hilbert space and diagonalize it, finding the ``raw" spectrum. As usual, we will employ a renormalization procedure to improve the precision. The necessary modifications are described in appendix \ref{sec:ren}. 

{\it Comparison with prior work:} The $\bZ_2$-broken phase of the $\phi^4$ model has been previously studied via a Hamiltonian truncation method in Ref.~\cite{Coser:2014lla}. There are many similarities between our works, and some differences. The main difference lies in the treatment of the zero mode (see also the discussion in \cite{Rychkov:2014eea}, section 4.5). Ref.~\cite{Coser:2014lla} compactifies the zero mode on a circle of large radius, and uses plane waves on this target space circle as the basis of trial wavefunctions. Instead, we resolve the zero mode dynamics and pick trial wavefunctions adapted to the quartic potential. Another difference is that they use conformal, massless, basis for the nonzero modes, while we use a massive basis. Matrix elements are easier to compute in the conformal basis, while a massive basis gives, we believe, a better initial approximation.

{Notice that} Ref.~\cite{Coser:2014lla} uses a different parametrization of the Hamiltonian, corresponding to a different normal-ordering prescription. Translation to our parametrization will be given in section \ref{sec:translation}.

\subsection{Varying the normal-ordering mass}

It turns out that in the regime we will be considering, the most important term inducing the interactions between $\mathcal{\hat{H}}_l$ and $\mathcal{\bar{H}}_l$ is the $\bar{V}_2$ term in \reef{eq:W}. This is because for the volumes that we will be able to consider, the localization of the $\phi_0$ wavefunctions near the potential minimum is not very sharp, and the coefficient of $\bar{V}_2$, viewed as a matrix in the space of the $\phi_0$ eigenstates, has significant matrix elements. The $\bar{V}_3$ and $\bar{V}_4$ terms will be suppressed at weak coupling.

Empirically, we concluded that the one-loop renormalization procedure, including the modifications to be described in appendix \ref{sec:ren}, is insufficient to fully describe the truncation effects arising from the big $\bar{V}_2$ term. Moreover, estimating the accuracy as the difference between the ``local" and ``subleading" renormalized answers was found inadequate in such a situation. Notice that the $V_2$ term renormalizes at quadratic order only the unit operator coefficient and this correction does not affect the spectrum of excitations \cite{Hogervorst:2014rta,Rychkov:2014eea} (this statement remains approximately true in the scheme with the separated zero mode discussed here). Ideally, to estimate the error one would have to compute the renormalization effects of cubic order in the problematic operator. Here we will resort to an interim alternative technique, which we now describe.\footnote{Another interesting possibility is to incorporate the coefficient of $\bar{V}_2$ into the mass of nonzero modes, making it $\phi_0$-dependent. This creates technical difficulties of its own and was not tried in this work.}

In the modified method as described in the previous section, the trial wavefunctions of the nonzero modes are taken to be those of the free massive boson of mass $M$, i.e.~the bare mass appearing in the Lagrangian. We will now consider the formalism in which one can vary the mass parameter $\mu$ of the trial wavefunctions. As in \cite{Lee:2000ac}, this will then be used to control the accuracy of our computations, since the exact spectrum should be independent of $\mu$. Apart from the accuracy issues, varying $\mu$ is also natural from the point of view of searching for an optimal zeroth order approximation to the ground state, in the spirit of variational methods.

So we rewrite the infinite-volume Hamiltonian (\ref{eq:H'}) by using the Coleman relations (\ref{eq:coleman}):
\begin{gather}
\label{eq:normordH}
H' =  \int d x N_\mu\bigl(\half\dot{\phi}^2+\half {\phi'}^2+( -{\textstyle\frac{1}{4}} M^2  + 6 g Z )\phi^2 + g \phi^4    + \Lambda_\mu\bigr)\,,
\\ \Lambda_\mu = \Lambda -{\textstyle\frac{1}{4}} M^2 Z + 3 g Z^2 + Y\,,
\end{gather}
where $Z=Z(M,\mu)$, $Y=Y(M,\mu)$ are defined in (\ref{eq:YZ}) with the replacement $M\to\mu, m\to M$. We then pass to finite volume as in section \ref{sec:dualtheory}:
\begin{gather}
\label{eq:normordH}
H'(L) = H_0 + [ -{\textstyle\frac{1}{4}} M^2 - \half \mu^2 + 6 (Z+\zeta)g ]V_2+ g V_4    + h_\mu\,,\\
h_\mu =  \Lambda_\mu L+ E_0+3 \zeta^2 g L 
+( -{\textstyle\frac{1}{4}} M^2 - \half \mu^2 + 6 g Z) \zeta L.
\end{gather}
where $H_0,V_2,V_4,E_0,\zeta$ are defined with respect to $\mu$. Finally, we separate the zero mode as in section \ref{sec:numres}. The final Hamiltonian has the form \reef{eq:ham0mode} where $\bar H_0=\bar H_0(L,\mu)$ while $\hat H$ and $W$ are given by:
\begin{gather}
\hat{H} = \frac{\NO{\pi_0^2}}{2L}+ L\left[ -{\textstyle \frac{1}{4}} M^2 + 6 (Z+\zeta) g \right]\NO{\phi_0^2} + L g\, \NO{\phi_0^4} +h_\mu \,,\\
W = \bigl [ 6 g\NO{\phi_0^2}-{\textstyle \frac{1}{4}} M^2  - \half \mu^2+ 6 (Z+\zeta) g\bigr] \bar V_2+ 4g \phi_0 \bar V_3+ g \bar V_4\,.
\end{gather}
This is the Hamiltonian which we use for numerical calculations, varying $\mu$ in the range $0.9$ - $1.1M$.
This will give an idea of the systematic error due to the truncation.

\subsection{Results}

From previous estimates, we know that the critical point lies at $g/m^2\approx 2.97(14)$ \cite{Rychkov:2014eea},\footnote{For more precise estimates by different methods see \cite{Wozar:2011gu,Milsted:2013rxa,Bosetti:2015lsa,Pelissetto:2015yha}.} which by making use of the Chang duality corresponds to $g/M^2 \approx 0.26$. Here we will limit ourselves to values $g/M^2 \le 0.2$, as beyond this value it appears difficult to reach the limit $L \to \infty$ and get a stable spectrum. $M$ will be set to $1$ throughout this section, unless stated otherwise.

We are now going to present the results for the two sectors of excitations of the theory. First, we will discuss the perturbative sector, which in the $L \to \infty$ consists of two decoupled towers of excitations around the two vacua with the opposite-sign VEV for the field. We will then turn to the non-perturbative sector of ``kink'' states which have topological charge, interpolating between the two vacua. Given the periodic boundary conditions imposed in our method, the kink sector will be studied here only indirectly, through the splitting of quasi-degenerate perturbative states in finite volume.

\subsubsection{Perturbative sector}
\label{sec:pertsector}

\begin{figure}[htb!]
\begin{center}
\includegraphics[scale=0.8]{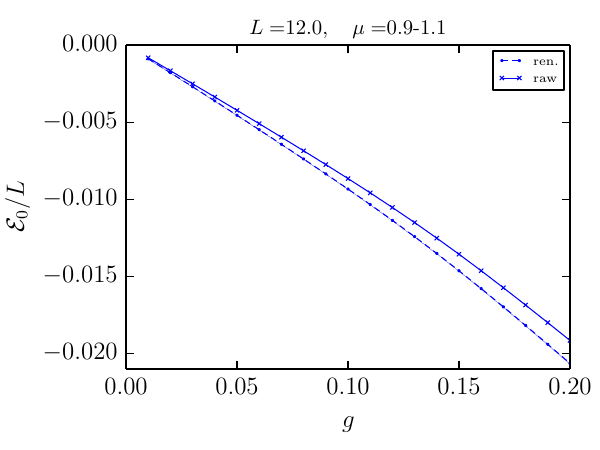} 
\includegraphics[scale=0.8]{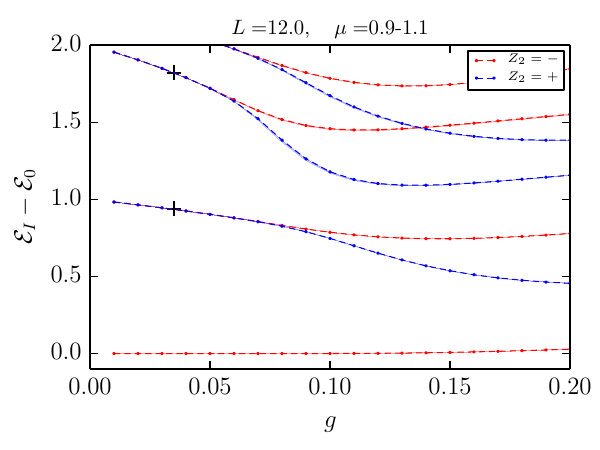}
\end{center}
\caption{The ground state energy density and the low-energy excitation spectrum as a function of $g$ for $L=12$; see the text. Results extracted from \cite{Coser:2014lla} are shown by crosses (whose size does not reflect the uncertainty), see section \ref{sec:translation}.}
\label{fig:vsG_L12}
\end{figure}

In figure \ref{fig:vsG_L12} we plot the ground state energy density and the low-energy excitation spectrum for $M=1$, $L=12$. For the ground state energy density we show both the ``raw'' and renormalized\footnote{In this section only local renormalization, in terminology of \cite{Rychkov:2014eea}, was used. Subleading nonlocal corrections were found to be totally negligible.} results, while for the spectrum only the renormalized results, because the raw/renormalized difference is negligible. As explained above, we don't think this difference gives a fair idea of the truncation error in the situation at hand. Instead, we estimate the error for the spectrum by varying the normal-ordering mass $\mu = 0.9$ - 1.1. In making these plots we fixed $s=4$, while the cutoff $\Emaxbar$ was chosen so that $\calH_l$ has dimension around $10000 - 15000$. We checked that increasing $s$ does not change the results significantly. 

We see that the first excited level is almost degenerate with the ground state. The splittings for the higher-energy levels are larger. This is because for the higher energy states it's easier to tunnel through the potential barrier separating the two infinite-volume vacua, which has a finite height for a finite $L$. 

\begin{figure}[htb!]
\begin{center}
\includegraphics[scale=0.8]{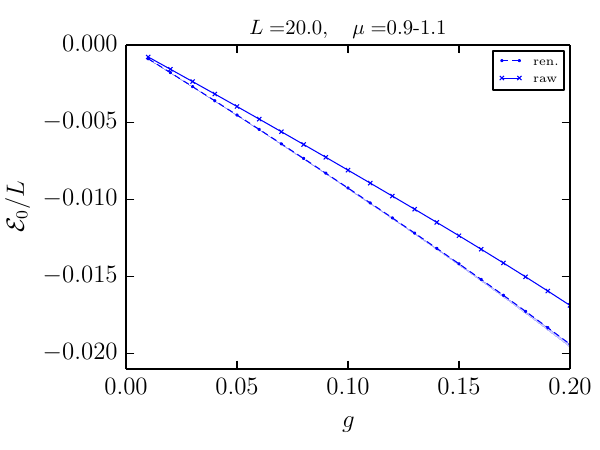} 
\includegraphics[scale=0.8]{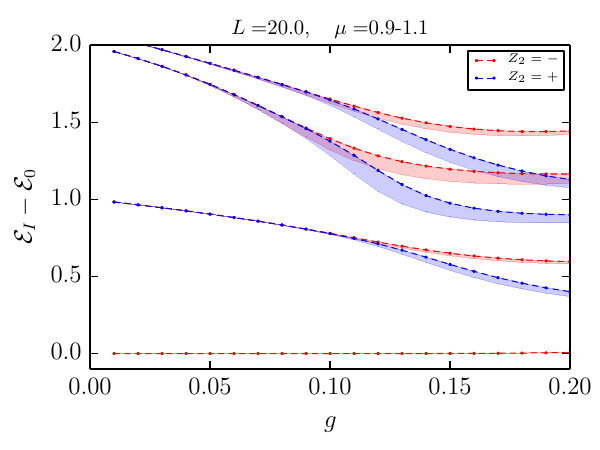}
\end{center}
\caption{Same as in figure \ref{fig:vsG_L20} but for $L=20$.}
\label{fig:vsG_L20}
\end{figure}

In figure \ref{fig:vsG_L20} we show the same plots for $L=20$. One can see that the energy splitting reduces but the truncation error increases (as one has to reduce $\Emaxbar$ in order to keep the total number of states the same).

\begin{figure}[htb!]
\begin{center}
\includegraphics[scale=0.8]{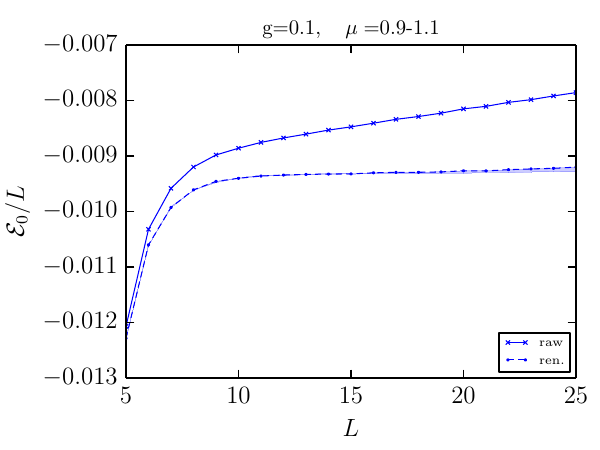} 
\includegraphics[scale=0.8]{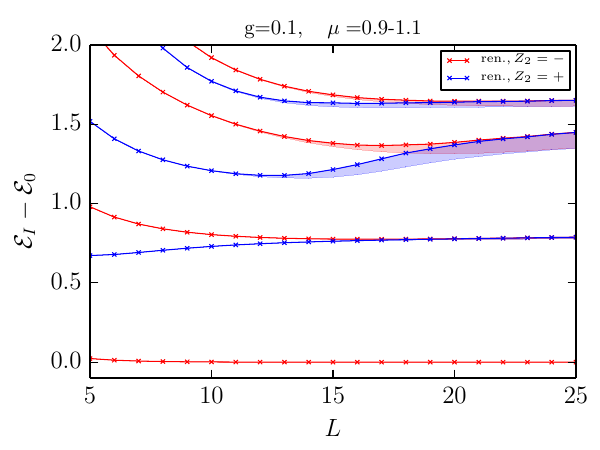}
\end{center}
\caption{Results for $g=0.1$ plotted as a function of $L$.}
\label{fig:vsL_G01}
\end{figure}

Finally, in figure \ref{fig:vsL_G01} we plot the vacuum energy density and the spectrum for $g=0.1$ as a function of $L$. One can see how the renormalization procedure is effective for the vacuum energy density, as its renormalized value reaches a constant for sufficiently large $L$, while its ``raw'' values does not. In the spectrum also the physical mass reaches a constant as expected.

Notice that for sufficiently small $g$ the interaction in the considered model is attractive (the cubic vertex squared attraction overcomes the quartic vertex repulsion) \cite{Caselle:2000yx,Coser:2014lla}. Therefore the second energy level pair in the spectrum in figure \ref{fig:vsL_G01} is expected to asymptote to $m_2 < 2 m_{\rm ph}$ (where $m_{\rm ph}$ is the single particle mass) as $L \to \infty$, i.e. it represents a bound state. The numerical results seem consistent with this expectation, although the precision is insufficient to extract $m_2$ accurately. In general, it is hard to extract the perturbative bound state mass from the infinite-volume limit, as the asymptotic convergence sets in at 
$L \approx (m^2_{\rm ph}-m_2^2/4)^{-1/2}$ \cite{Luscher:1985dn}, 
which diverges as $g \to 0$.

In Appendix \ref{sec:pert} we compare the numerical results for $\Lambda$ and $m_{\rm ph}$ with the predictions from perturbation theory, showing very good agreement at small couplings.

\begin{figure}[htb!]
\begin{center}
\includegraphics[scale=0.8]{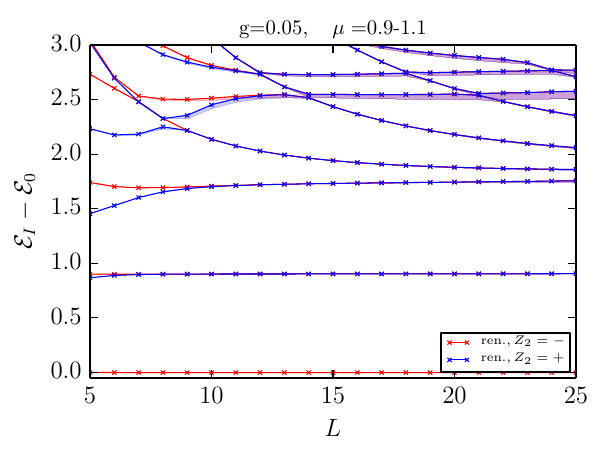} 
\end{center}
\caption{Same as in the right-hand figure \ref{fig:vsL_G01} but for $g=0.05$.}
\label{fig:vsL_G005}
\end{figure}

It is also interesting to analyze the higher-energy states in the spectrum. In figure \ref{fig:vsL_G005} we redo the previous plot for $g=0.05$, including a few more eigenvalues. Above the stable particle mass and the bound state, one can see the multiparticle states whose energy depends on $L$ according to the dispersion relations in finite volume.\footnote{See e.g.~the discussion in \cite{Coser:2014lla}, appendix B.} Furthermore, the horizontal line with energy $\approx 2.5 < 3 m_{\rm ph}$  represents a resonance. Due to the non-integrability of the theory, that state is not stable, as its energy is larger than  $2 m_{\rm ph}$. Indeed, the horizontal line does not cross the multiparticle states as could seem at first glance, thanks to the phenomenon of avoided crossing. See \cite{Delfino:1996xp} for a discussion of how resonances should appear in the finite volume spectrum. 

\subsubsection{Non-perturbative sector}
\label{sec:NP}
As already mentioned, in finite volume non-perturbative effects lift the spectrum degeneracy both for the ground state and for all the excited states. 
For small coupling, these effects can be interpreted as tunneling due to the semiclassical field configurations interpolating between the two vacua (``kinks"). The splitting depends on the mass of the kink. Here we will need the semiclassical prediction for the splitting of the first two energy levels (the ground state, which lives in the $\bZ_2$ even sector, and the $\bZ_2$ odd state just above it). Including the leading semiclassical results and the one-loop determinant fluctuations around it, the splitting for small $g/M^2$ is given by (see appendix \ref{sec:energysplitting}):
\begin{equation}
\label{eq:splitting}
\Delta \calE = \mathcal{E}_1-\mathcal{E}_0  \approx \sqrt{\frac{M^3}{6 \pi g L}} e^{-L M_{\rm kink}-f(ML)} \,, \qquad M_{\rm kink} = \frac{M^3}{12 g} + M \left( \frac{1}{4 \sqrt{3}} - \frac{3}{2 \pi}\right)\,,
\end{equation}
where $M_{\rm kink}$ is the kink mass in the one-loop approximation, first computed in \cite{Dashen:1974cj}. Corrections are suppressed by $g/M^2$ and by $1/(LM_{\rm kink})$. The function $f(x)$, given in \reef{eq:f(x)}, approaches zero exponentially fast for $LM\gg 1$.

\begin{figure}[htb!]
\begin{center}
\includegraphics[scale=1]{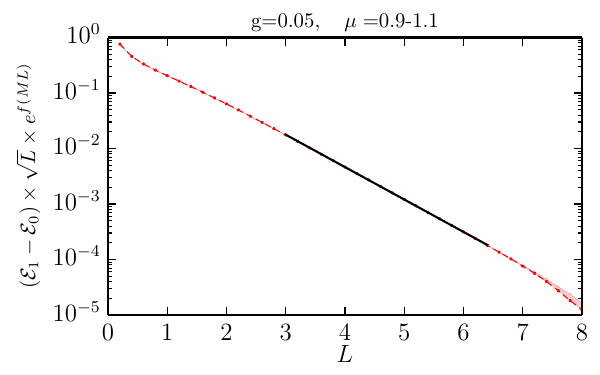}
\end{center}
\caption{Ground state splitting as a function of $L$ for $g=0.05$; see the text.}
\label{fig:fitkink_G005}
\end{figure}

Our numerical method allows to extract $\Delta \calE$ with high precision and to compare with this formula. In figure \ref{fig:fitkink_G005} we present as an example the renormalized numerical results\footnote{The difference between ``raw'' and renormalized is negligible in the present analysis.} for $M=1$, $g=0.05$. We used $s = 5$, checking that its increase does not change significantly the numerics, while $\Emaxbar$ was fixed such as to have a basis dimension $\sim 10000$ for each $L$.
We plot $ \sqrt{L}\,e^{f(ML)}\Delta \calE$ as a function of $L$ in logarithmic scale in order to observe a linear trend, as expected from (\ref{eq:splitting}), and perform a fit in a region chosen by eye such that the data look close to a straight line:
\begin{equation}
\log \bigl[\sqrt{L}\,e^{f(ML)}\Delta \calE] \approx \alpha - M_{*} L\,.
\end{equation}
The value of $L$ must be not too low so that the exponential law decay sets in, and not too high otherwise $\Delta \calE$ becomes smaller than the precision of our method. We then compare the fitted values of $\alpha$ and $M_*$ with the expectations from \reef{eq:splitting}.

\begin{figure}[htb!]
\begin{center}
\includegraphics[scale=1]{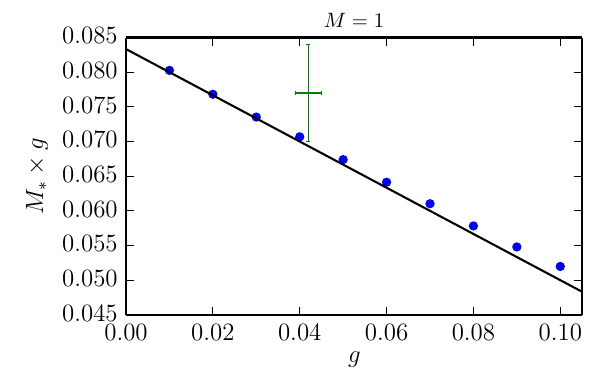}
\end{center}
\caption{Comparison between the fitted and the theoretically predicted value of the kink mass; see the text. The green cross represents, with error bars, a result from \cite{Coser:2014lla} as discussed in section \ref{sec:translation}.}
\label{fig:fitkink}
\end{figure}

We carried out this analysis for several values of the coupling between $0.01$ and $0.1$, finding both $\alpha$ and $M_*$ very close to the expected values. The comparison of $M_*$ with $M_{\rm kink}$ is plotted in figure \ref{fig:fitkink} as a function of $g$. It turns out that in the range of points where the fit is made $f(ML)$ is very small and does not influence the fit, except a little for the smallest considered values of $g$. On the other hand including $\sqrt{L}$ is crucial for reaching the agreement. One can see that the accord with the semiclassical prediction $M_{\rm kink}$ (black line) is very good.

\subsubsection{Comparison to Ref.~[3]}
\label{sec:translation}

For comparison we included in figures \ref{fig:vsG_L12},\ref{fig:fitkink} a few data points extracted from \cite{Coser:2014lla}. Ref.~\cite{Coser:2014lla} parametrizes the theory by two couplings $G_2,G_4$ which they denote $g_2,g_4$; we capitalized to avoid confusion with our notation in other parts of this paper. Their couplings are not identical to ours; because of the different field normalization $g=2\pi G_4$. More importantly, their $\phi^4$ operator is normal-ordered differently, by subtracting the normal-ordering constants for all nonzero massless modes in finite volume $L$(= their $R$). Going to our normal ordering prescription (in infinite volume), 
\beq
\NO{\phi^4}_{\rm their}\to N_m(\phi^4)-C(mL)N_m(\phi^2)+const.,\qquad C(mL)= -(3/\pi)\log[e^\gamma mL/(4\pi)]\,,
\eeq 
where $\gamma$ is the Euler-Mascheroni constant. We don't pay attention to the ground state energy renormalization here. To put their Hamiltonian into the canonical form \reef{eq:H} (resp.~\reef{eq:H'0}) one has to solve the two equations
\beq
G_2-2gC(mL)=m^2\quad (\text{resp.}\quad
G_2-2gC(ML)=-M^2/2)
\eeq
for $m$ or $M$ respectively. Keeping $G_{2,4}$ fixed and varying $L$ thus induces a logarithmic variation of the infinite-volume mass parameters. Although for the small quartic couplings considered in \cite{Coser:2014lla} this variation is not huge (order $10\%$), it may be problematic for extracting the spectrum by approaching the large $L$ limit. It would seem more appropriate to vary $G_2$ with $L$ while keeping $m$ or $M$ fixed.

The two data points (crosses) in figure \ref{fig:vsG_L12} were extracted from figure 10(b,d) of \cite{Coser:2014lla}, where $G_2=-0.1$, $G_4=1.2\times 10^{-3}$. This corresponds to $g/M^2\approx 0.035$ at $ML=12$. The agreement between their and our results is good. Their determination of the kink mass for the same $G_{2,4}$ is shown in figure \ref{fig:fitkink}. Here $g/M^2=0.042(3)$, varying within the range of $L$ used in their fit. The large error bars on $M_{\rm kink}$ may be due to this variation. Also, they did not consider the pre-exponential factor in (\ref{eq:splitting}).

A remark is in order concerning the discussion in \cite{Mussardo:2006iv,Coser:2014lla}, which views the particles in the topologically trivial sector as bound states of kinks. A semiclassical prediction is given for their masses (\cite{Coser:2014lla}, (28)):
\beq 
m_{{\rm sc},n}=2M_{\rm kink} \sin(n \pi\xi/2),\quad n=1\ldots[1/\xi],
\label{eq:semi}
\eeq
where 
\beq
\xi=M/(\pi M_{\rm kink})
\eeq
in our notation. The lightest mass $m_{{\rm sc},1}$ has to be identified with our $m_{\rm ph}$, while the second $m_{{\rm sc},2}$ with the bound state mass $m_2$ discussed in section \ref{sec:pertsector}. The other masses correspond not to stable particles but to resonances in a non-integrable theory like the one we are considering. The total number of particles is predicted to be $[1/\xi]$. 

The semiclassical prediction is valid for $\xi\ll1$, but if one could extrapolate it to $\xi=O(1)$ one would naively predict that for $\xi>1$ the topologically trivial sector would be devoid of particles. This would be analogous to the phase of the sine-Gordon model for $4\pi<\beta^2<8\pi$. Of course it's far from clear if such an extrapolation is trustworthy.


From the kink mass formula \reef{eq:splitting} we have $\xi=1$ for $g/M^2\approx 0.12$, just outside the region that we explored, and well below the critical point at $g_c/M^2\approx 0.26$. It will be interesting to study this range in the future. One minimalistic possibility is that the topologically trivial particles disappear only at the critical point. Indeed, its neighborhood is described by the thermally perturbed 2d Ising model CFT, which is free massive Majorana fermion theory. In the low-temperature phase, the fermionic excitations are naturally identified with the kink states interpolating between the two vacua. There are no bound states since the fermions are free.\footnote{{\bf Note added:} This `minimalistic possibility' can now be ruled out, in favor of the original scenario of  \cite{Mussardo:2006iv}, based on the very recent results of \cite{Bajnok:2015bgw}. As this paper shows, the second lightest particle $m_2$ in the topologically trivial spectrum becomes unstable with respect to decay into two kinks for $g/M^2\gtrsim 0.075$, while for the lightest particle $m_{\rm ph}$ this happens for $g/M^2\gtrsim 0.125$. The possibility of the first of these decays could be observed already from our mass plots in Figs.~\ref{fig:vsG_L20},\ref{fig:fitkink}.}

\section{Conclusions}
\label{sec:conclusions}

In this paper we followed up on our earlier study \cite{Rychkov:2014eea} of the Hamiltonian truncation technique applied to the $\phi^4$ theory in two dimensions. The main results derived in this work can be summarized as follows:
\begin{itemize}
\item According to an exact duality, reviewed in section \ref{sec:chang}, the theory under consideration can be expressed via two different Lagrangian formulations. We proved that, even at strong coupling, the Hamiltonian truncation method correctly predicts the same low-energy spectrum of excitations in the two cases, despite the fact that they look totally different at the zeroth order. We regard this as a non-trivial check of the method.

\item We showed how to modify the method in order improve its accuracy in the spontaneously broken phase. 
We found very good agreement with the predictions from perturbation theory and semiclassics in the perturbative and non-perturbative sectors. To approach the critical region as in \cite{Rychkov:2014eea} will require further improvements of the method.\footnote{{\bf Note added:} Rapid progress in this direction should be possible thanks to the technical and conceptual improvements discussed in \cite{Bajnok:2015bgw,Elias-Miro:2015bqk}, which appeared a few weeks after our work.}
\end{itemize}

We continue to believe that the potential of ``exact diagonalization'' techniques, among which we have implemented a particular realization in the present work, is very large and has to be explored further. 
Some other representative applications to non-integrable theories to be found in the literature are \cite{Delfino:1996xp,Bajnok:2000ar,Fonseca:2006au,Caux:2012nk,Coser:2014lla,2014arXiv1407.7167B,Konik:2015bia,Lepori:2009ip,Lencses:2015bpa} in $d=2$. In $d>2$ the only work is \cite{Hogervorst:2014rta}. 

In the future it would be interesting to extend the present analysis, for instance by studying the topological spectrum of kink-states directly,\footnote{{\bf Note added:} This has just been achieved in \cite{Bajnok:2015bgw}. The authors use a different truncation scheme and diagonalization routine, and they are able to calculate the kink mass up to $g \sim 0.2$ (in our conventions).} or consider more complicated theories involving scalar-fermion interactions, which should not be too technically challenging.\footnote{See \cite{Brooks:1983sb} for early work.} In the long term, in order to solve numerically higher dimensional theories, it will be necessary at the very least to refine the renormalization technique, as the RG flow becomes more weakly relevant.\footnote{{\bf Note added:} See \cite{Elias-Miro:2015bqk} for recent progress towards the calculation of higher order renormalization coefficients.} The hope is that exact diagonalization techniques can evolve into computationally efficient tools to address difficult problems in quantum field theory.
\section*{Acknowledgements}

We are grateful to Giuseppe Mussardo for the comments of the draft. This research was partly supported by the National Centre of Competence in Research SwissMAP, funded by the Swiss National Science Foundation. The work of L.V. is supported by the Swiss National Science Foundation under grant 200020-150060.

\appendix

\section{Renormalization in the $\bZ_2$-broken phase}
\label{sec:ren}

We invite the reader to go first through the explanation of the renormalization procedures for the $\bZ_2$-symmetric phase presented in detail in \cite{Rychkov:2014eea}, as the notation and logic below are closely inherited from that discussion.

Let us start from the full eigenvalue problem:
\begin{equation}
\label{eq:eigeqH}
H.c = \mathcal{E} c\,.
\end{equation}
The full Hilbert space can be split into ``low'' energy and ``high'' energy subspaces:
\begin{gather}
H_l = \mathcal{\hat{H}}_l \otimes \mathcal{\bar{H}}_l\,, \\
H_h = (\mathcal{\hat{H}}_h \otimes \mathcal{\bar{H}}_l )\oplus (\mathcal{\hat{H}}_l \otimes \mathcal{\bar{H}}_h )\oplus (\mathcal{\hat{H}}_h \otimes \mathcal{\bar{H}}_h)\,.
\end{gather}
Accordingly, (\ref{eq:eigeqH}) can be projected onto these subspaces:
\begin{gather}
H_{ll}.c_l + H_{lh}.c_h= \mathcal{E} c_l\,, \\
H_{hl}.c_l + H_{hh}.c_h= \mathcal{E} c_h\,.
\end{gather}
``Integrating out'' $c_h$ via the second equation, we get:
\begin{gather}
(H_{ll} + \Delta H) c_l = \mathcal{E} c_l\,, \\
\Delta H = - H_{l h} \frac{1}{H_{h h}-\mathcal{E}} H_{h l} =  - W_{l h} \frac{1}{H_{h h}-\mathcal{E}} W_{h l}\,, \label{eq:deltah}
\end{gather}
where we used that in the Hamiltionian (\ref{eq:ham0mode}) only $W$ will mix the low and high subspaces. At leading order one can neglect $W$ in the denominator, which gives:
\begin{equation}
\label{eq:deltahLO}
\Delta H \approx - W_{l h} \frac{1}{\hat{H} +\bar{H}_0 -\mathcal{E}} W_{h l} = - \sum_{i \in \mathcal{H}_h} \frac{1}{E_i - \mathcal{E}} P_l W \ket{i}\bra{i} W P_l
\end{equation}
where a summation over all the states in $\mathcal{H}_h$ appears. 

It turns out that the effect induced by the truncation of $\hat{\calH}$ is less significant than for $\bar{\calH}$. It's also less expensive to control. We found that fixing the corresponding cutoff $s$ to 4 or 5 basically stabilizes the results. For this reason we will only take into account the renormalization effect coming from the nonzero field modes. This means that we will restrict the sum in \reef{eq:deltahLO} to go only over the $\mathcal{\hat{H}}_l \otimes \mathcal{\bar{H}}_h$ part of $\calH_h$. Therefore, we approximate:
\begin{equation}
\begin{split}
\Delta H  & \approx - \sum_{\psi_\alpha \in \hat{\mathcal{H}}_l} \sum_{k \in \bar{\mathcal{H}}_h} \frac{1}{\hat{\mathcal{E}}_\alpha + \bar{E}_k- \mathcal{E}} W \ket{\psi_\alpha, k}\bra{\psi_\alpha, k} W 
\end{split}
\end{equation}
where we dropped the projectors $P_l$ to avoid cluttering.
The potential matrix $W$ can be schematically written as:
\begin{equation}
W = \sum_{a=2,3,4} \hat{m}_a \otimes \bar{V}_a\,,
\end{equation}
where $\hat{m}_a$ and $\bar{V}_a$ are matrices in the $\hat{\mathcal{H}}$ and $\bar{\mathcal{H}}$, respectively.
Therefore: 
\begin{equation}
\label{eq:deltah1}
\Delta H \approx - \sum_{a, b}  \sum_{\psi_\alpha \in \hat{\mathcal{H}}_l}  \Bigl(\hat{m}_a \ket{\psi_\alpha} \bra{\psi_\alpha}\hat{m}_b \Bigr) \otimes  \Biggl( \sum_{k \in \bar{\mathcal{H}}_h} \frac{1}{\hat{\mathcal{E}}_\alpha+ \bar{E}_k- \mathcal{E}} \bar{V}_a\ket{k}\bra{k}\bar{V}_b \Biggr) \,.
\end{equation}
The matrix elements $\Bigl(\hat{m}_a \ket{\psi_\alpha} \bra{\psi_\alpha}\hat{m}_b \Bigr)$ can be computed explicitly, while the second factor in (\ref{eq:deltah1}) is evaluated with the same technique developed in \cite{Rychkov:2014eea}:
\begin{gather}
 \sum_{k \in \bar{\mathcal{H}}_h} \frac{1}{\hat{\mathcal{E}}_\alpha + \bar{E}_k- \mathcal{E}} \bar{V}_a\ket{k}\bra{k}\bar{V}_b = \int_{\Emaxbar}^\infty dE \frac{1}{\hat{\mathcal{E}}_\alpha + E- \mathcal{E}} M^{a b}(E)\,, \\
 M^{a b}(E)_{i j} \, dE \equiv \sum_{k: E \le \bar{E}_k \le E+dE} (\bar{V}_a)_{i k} (\bar{V}_b)_{k j}\,,
\end{gather}
where the matrix elements $M^{a b}_{i j}$ can be related to the non-analytic behavior of two-point functions of the potential operators:
\begin{gather}
C^{a b}(\tau)_{i j} = \braket{i}{\bar{V}_a(\tau/2) \bar{V}_b(-\tau/2)}{j} = \int_0^\infty e^{-\left[E-(E_i+E_j)/2 \right]} M^{a b}_{i j}(E)\,, \\
\bar{V}_a(\tau) \equiv e^{H_0 \tau} \bar{V}_a e^{-H_0 \tau}\,.
\end{gather}
The quantities $C^{a b}(\tau)_{i j}$ are computed as in \cite{Rychkov:2014eea} using the Wick theorem. The only difference is that the boson two point function $\bar{G}(\rho)$ in the present case does not include the contribution from the zero mode:
\begin{equation}
\bar{G}(\rho) = G(\rho) - \frac{1}{2 L M}\,.
\end{equation}

\section{Perturbation theory checks}
\label{sec:pert}

We computed the first few perturbative corrections to the ground state energy density $\Lambda$ and the physical particle mass $m_{\rm ph}$ for the potential density:
\begin{equation}
\mathcal{V} = \half M^2 N_M(\phi^2) + g_3 N_M(\phi^3) + g_4 N_M(\phi^4)
\end{equation}
The symmetric double-well case of Eq.~(\ref{eq:shiftedpotential}) can be recovered by setting $g_4 =g$, $g_3 = \sqrt{2 g_4} M$, but we will keep the couplings independent for the sake of generality.

For comparison with numerics, we will need results for $\Delta m^2=m_{\rm ph}^2-M^2$ and $\Lambda$ up to the second order in $g$. In terms of $g_3,g_4$, we need to include all diagrams up to order $O(g_4^2)$, $O(g_3^2)$, $O(g_3^2 g_4)$ and $O(g_3^4)$. The results are\footnote{We do not report explicitly the symmetry factors for the diagrams. Most of them were evaluated numerically by Monte Carlo integration using coordinate space propagators. We did not invest much effort in analytic results.} ($\bar{g}_3\equiv g_3/M^2$, $\bar{g}_4\equiv g_4/M^2$):
\begin{align}
\Lambda / M^2 &= 
\vcenter{\hbox{\includegraphics[trim= .5cm 0cm .5cm 0cm, clip=true, scale=0.3]{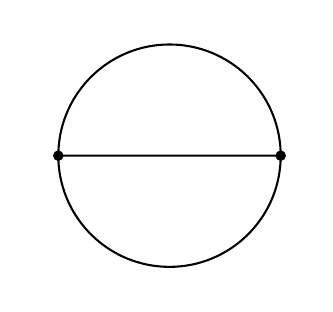}}} +
\vcenter{\hbox{\includegraphics[trim= .5cm 0cm .5cm 0cm, clip=true, scale=0.3]{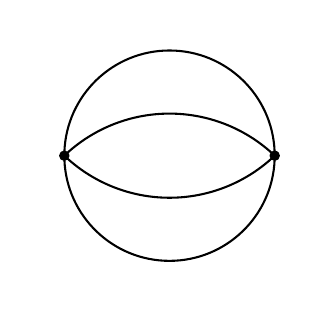}}} +
\vcenter{\hbox{\includegraphics[trim= 6.5cm 11cm 6.5cm 3cm, clip=true, scale=0.3]{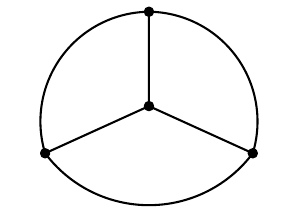}}} +
\vcenter{\hbox{\includegraphics[trim= 1cm 0cm 1cm 0cm, clip=true, scale=0.3]{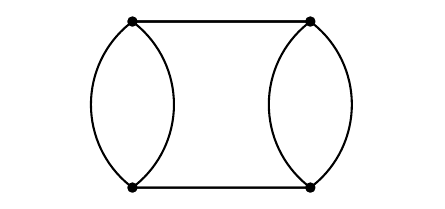}}} +
\vcenter{\hbox{\includegraphics[trim= 3cm 0cm 3cm 0.5cm, clip=true, scale=0.4]{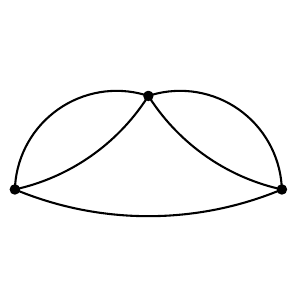}}}
+ \,\ldots
\\ &= -0.0445289 \bar{g}_3^2 -\frac{21 \zeta(3)}{16 \pi^3}\bar{g}_4^2 - (0.0109030(51)+0.026854(32))\bar{g}_3^4 + 0.0799586(41) \bar{g}_3^2 \bar{g}_4 + \ldots\nn
\end{align}
and 
\begin{align}
\label{eq:masscorrwrong}
\Delta m^2/M^2&=
\vcenter{\hbox{\includegraphics[trim= .5cm 1cm .5cm 1cm, clip=true, scale=0.33]{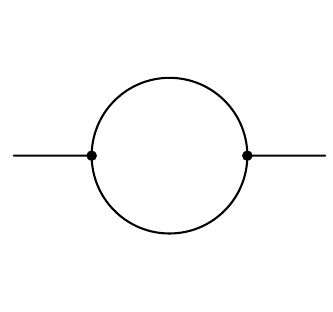}}} +
\vcenter{\hbox{\includegraphics[trim= .5cm 1cm .5cm 1cm, clip=true, scale=0.4]{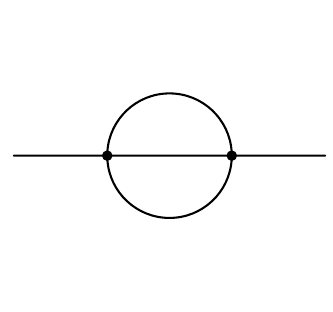}}} +
\vcenter{\hbox{\includegraphics[trim= .5cm 1cm .5cm 1cm, clip=true, scale=0.4]{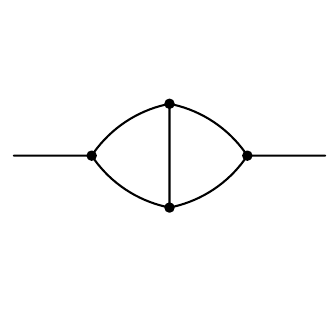}}} +
\vcenter{\hbox{\includegraphics[trim= .5cm 1cm .5cm 1cm, clip=true, scale=0.45]{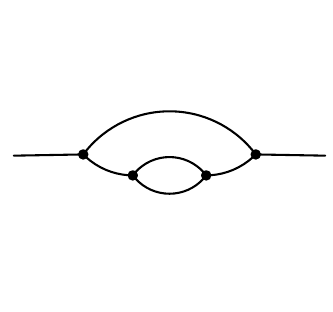}}} \nn\\
&\hspace{2cm}
+
\vcenter{\hbox{\includegraphics[trim= 1cm 1cm 1cm 1cm, clip=true, scale=0.45]{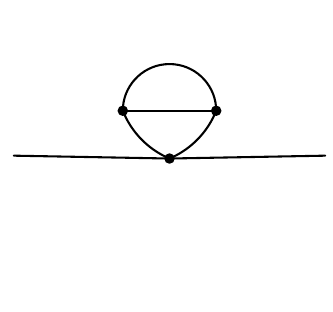}}} +
\vcenter{\hbox{\includegraphics[trim= .5cm 1cm .5cm 1cm, clip=true, scale=0.45]{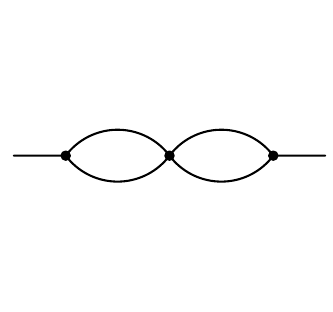}}} +
\vcenter{\hbox{\includegraphics[trim= .5cm 1cm .5cm 1cm, clip=true, scale=0.45]{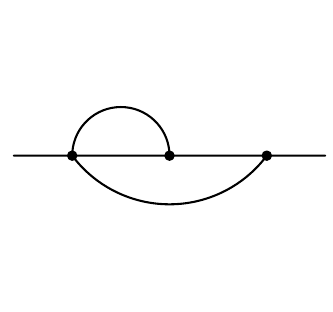}}} +
\,\ldots\nn
\\ &= -\sqrt{3} \bar{g}_3^2 -1.5 \bar{g}_4^2 - (2.2492(37)+2.8020(70))\bar{g}_3^4\nn\\[5pt]
&\hspace{2cm}+\ (1.06864(15)+1.9998(10)+5.50025(91))\bar{g}_3^2\bar{g}_4 + \ldots
\end{align}

\begin{figure}[h!]
\begin{center}
\includegraphics[scale=0.85]{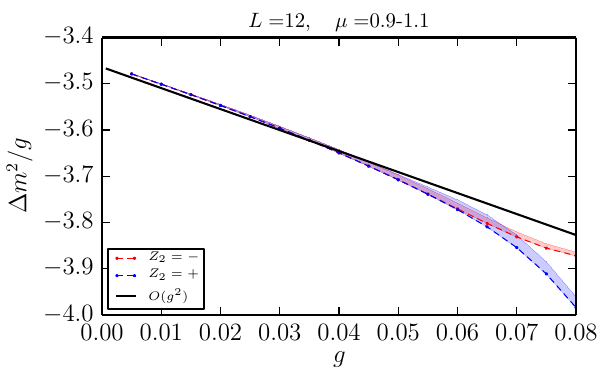}
\includegraphics[scale=0.85]{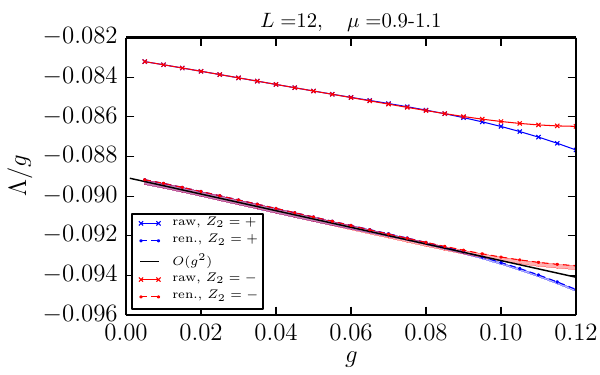} 
\end{center}
\caption{Comparing perturbative and numerical predictions; see the text.}
\label{fig:phi4pert}
\end{figure}

In figure \ref{fig:phi4pert} we compare the above predictions for $g_4=g$, $g_3 = \sqrt{2 g_4} M$ with the numerical spectra obtained with our method for $M=1$, $L=12$. We use the zero-mode cutoff $s=4$ and adjust $\Emaxbar$ so that the basis dimension is $\sim 10000$ in each sector.

In the left plot we show the renormalized results for $\Delta m^2$, computed both in the $\bZ_2$-even and $\bZ_2$-odd spectra, with an error estimate given by variation of the normal ordering mass. We observe a reasonably good agreement for $g\lesssim 0.04$.\footnote{We haven't investigated the reasons behind a small residual deviation visible in this region. One possible reason is that we may be underestimating the renormalization corrections by including contributions only from the $\mathcal{\hat{H}}_l \otimes \mathcal{\bar{H}}_h$ part of the high energy Hilbert space. See appendix \ref{sec:ren}.} For larger $g$, the deviation may be attributed to higher-order perturbative effects and to the finite-volume splitting affecting the numerics.

In the right plot we show instead both the ``raw'' and renormalized results for the ground state energy density, extracted from both $\bZ_2$-even and $\bZ_2$-odd spectra. Again, an error estimate for the renormalized values is attributed by varying the normal ordering mass. We see a perfect agreement with the perturbative prediction until the finite-volume splitting between the eigenvalues kicks in. 

\begin{figure}[h!]
\begin{center}
\includegraphics[scale=0.85]{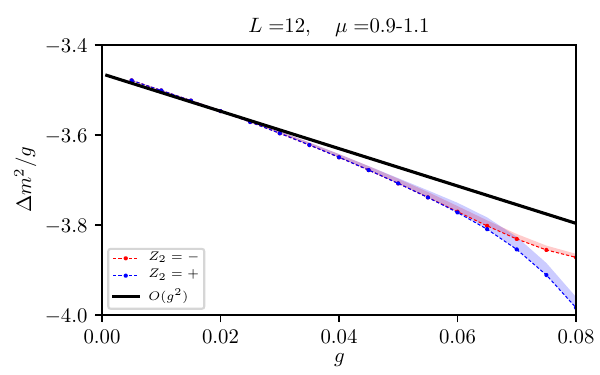}
\end{center}
\caption{Comparing perturbative and numerical predictions for the mass gap, 
with corrected perturbative coefficients; see the text.}
\label{fig:phi4pertnew}
\end{figure}

\textbf{Note added} (April 2019) Eq. 
\eqref{eq:masscorrwrong} contains several mistakes. First, errors from Monte Carlo integration in diagrams of order $g_3^4$ and $g_3^2 g_4$ are underestimated. Second, we missed two tadpole diagrams (one at order $g_3^4$ and one at order $g_3^2 g_4$). 
Third, the correct formula should include wavefunction-renormalization effects, which at the given order come 
from the $p^2$ dependence of the first diagram in  \eqref{eq:masscorrwrong} 
and further correct the $g_3^4$ coefficient. Taking into account all these effects,
and substituting $g_3=\sqrt{2 g_4}$, the correct formula for the mass correction 
is (see \cite{Serone:2019szm}, Eq. (4.12))
$\Delta m^2/M^2 = - 2 \sqrt{3} g_4 - 4.1529(18) g_4^2 + \ldots$. 
With this formula, the left plot in Fig.\ref{fig:phi4pert} is replaced by  
Fig.\ref{fig:phi4pertnew}. We thank Gabriele Spada for pointing out that we missed the tadpole diagrams.

\section{Ground state splitting}
\label{sec:energysplitting}

We will review here the derivation of Eq.~\reef{eq:splitting}.\footnote{See \cite{Vainshtein:1981wh} for a pedagogical discussion in quantum mechanics, and \cite{Osborn:1978qg} for an analogous computation for the partition function at finite temperature, which can be interpreted as a computation of the coefficient $\kappa$ in \reef{eq:amplitude3}.} We start from the Euclidean action:
\begin{equation}
S = \int d^2 x \left[ \half (\partial \phi)^2 + g \left(\phi^2 - c^2\right)^2 \right]\,,
\end{equation}
which entails the perturbative particle mass $m^2 = 8 g c^2$. The normal ordering prescription for renormalization adopted in this work is equivalent to the mass renormalization:\footnote{Let us neglect the cosmological constant renormalization as it does not affect the energy splitting.}
\beq
S \to S - \frac{\delta m^2}{4}\int d^2 x\, \phi^2\,,\qquad  \delta m^2 = \frac{6 g}{\pi} \int_{-\infty}^{\infty} dk \frac{1}{\sqrt{k^2+m^2}}\,.
\eeq

We will compute the matrix elements:
\begin{equation}
\label{eq:amplitude}
\calA_\pm=\braket{\phi = \pm c}{e^{-H \tau_0}}{\phi = c} = \calN \int \mathcal{D}\phi\,\, e^{-S[\phi]}\,,
\end{equation}
where the path integral in the r.h.s. is defined with the boundary conditions $\phi(x,\tau_0/2) = \pm c$, $\phi(x,-\tau_0/2) = c$. The path integral measure normalization factor $\calN$ will be fixed below. The results for the matrix elements will then be translated into the energy splitting.

Consider first the transition amplitude from $c$ to $-c$ in the one-instanton approximation. The instanton takes the form:
\begin{equation}
\phi_0(x,\tau) = c \tanh \frac{m (\tau-\tau_c)}{2}\,,
\end{equation}
where the center $\tau_c$ is arbitrary. This solution has action:
\begin{equation}
\label{eq:action}
S_0 + \delta S_0 = L \frac{m^3}{12 g} - L\, \delta m^2 \frac{c^2}{4} \int_{-\infty}^{\infty} d \tau \left(\tanh^2 \frac{m \tau}{2} -1 \right) = L \frac{m^3}{12 g} + L \frac{\delta m^2}{8 g} m\,,
\end{equation}
where we included the contribution due to the mass counterterm. We need $S_0\gg 1$ for the validity of the semiclassical approximation.

At the one-loop order, (\ref{eq:amplitude}) can be approximated by:
\begin{equation}
\calN \int \mathcal{D}\phi\,\, e^{-S[\phi]} \approx e^{- S_0} \calN \int \mathcal{D} \eta \,\, e^{- \int \eta \frac{\delta^2 S}{\delta \phi^2} \eta }\,.
\end{equation}
Taking into account the presence of the zero mode of the quadratic fluctuation operator $\frac{\delta^2 S}{\delta \phi^2}$ due to the invariance of $S$ under a shift of $\tau_c$, this results in:
\begin{equation}
\label{eq:amplitude2}
\sqrt{\frac{S_0}{2 \pi}}  e^{-S_0} \calN \left[ \det\nolimits'{\left(  -\square + V'' \right)}\right]^{-1/2} \tau_0\,,
\end{equation}
where the prime indicates that the zero mode has been removed from the determinant, and we replaced the integral over the zero mode with \cite{Vainshtein:1981wh}:
\begin{equation}
\int d c_0 = \sqrt{\frac{S_0}{2 \pi}} \int_{-\tau_0/2}^{\tau_0/2} d \tau\,.
\end{equation}

To fix $\calN$, consider the $0\to0$ transition amplitude in the free massive theory, given simply by
\begin{equation}
\calA_0=\braket{\phi = 0}{e^{-H_0 \tau_0}}{\phi=0}= \calN \left[ \det{\left(-\square + m^2\right)}\right]^{-1/2}\,.
\end{equation}
In the ratio of the two amplitudes the normalization factor cancels:
\begin{equation}
\label{eq:amplitude3}
\calA_-/\calA_0\approx \kappa\tau_0\,,
\qquad \kappa=\sqrt{\frac{S_0}{2 \pi}} e^{-S_0} \left[\frac{\det\nolimits'{\left( -\square + V'' \right)}}{m^{-2}\det{\left(-\square + m^2\right)}} \right]^{-1/2} m \,. 
\end{equation}

Now, let us calculate the determinant ratio. We need to solve the eigenvalue equation:
\begin{equation}
\left(- \frac{d^2}{d x^2} - \frac{d^2}{d \tau^2} + 12 g \phi_0^2 - 4 g c^2 \right) \psi = \biggl(- \frac{d^2}{d x^2} - \frac{d^2}{d \tau^2} +m^2 -\frac{3}{2}m^2 \frac{1}{\cosh^2 \frac{m \tau}{2}}  \biggr) \psi = \epsilon \psi\,.
\end{equation}
The eigenstates are of the form $\psi(x,\tau)= e^{- i k_n x} \psi_n(\tau)$, where $k_n = \frac{2 \pi n}{L}$ due to periodic boundary condition on the cylinder, and
\begin{equation}
\label{eq:eigeq}
\biggl(- \frac{d^2}{d \tau^2} +\omega_n^2 -\frac{3}{2}m^2 \frac{1}{\cosh^2 \frac{m \tau}{2}}  \biggr) \psi_n = \epsilon_n \psi_n\,,
\end{equation}
where we defined $\omega_n^2 \equiv k_n^2+m^2$. The eigenvalues of (\ref{eq:eigeq}) comprise two bound states:
\begin{equation}
\epsilon_{n,0} = k_n^2 , \qquad \epsilon_{n,1} = \frac{3}{4}m^2 + k_n^2
\end{equation}
and a continuum (for infinite $\tau_0$) of states with $\epsilon_n \ge \omega_n^2$ \cite{feshbach},  which can be parametrized by the ``momentum'':
\begin{equation}
p = \sqrt{\epsilon_n} - \omega_n\ge 0\,.
\end{equation}
We consider $\tau_0\gg m^{-1}$ large but finite (but not too large---see below). Imposing the boundary conditions $\psi_n(\pm\tau_0/2)=0$, the $p$'s take discrete values:
\begin{equation}
p \tau_0 - \delta_p = \pi l = \tilde{p}_l \tau_0 \,, \qquad l = 0, 1, \ldots\,.
\end{equation}
where the $\tilde{p}_l$ represent the eigenvalues in the free theory, and the phase shift is \cite{feshbach}:
\begin{equation}
\delta_p = -2 \pi + 2 \arctan \frac{2 p}{m} + 2 \arctan \frac{p}{m}
\end{equation}
Here the $-2 \pi$ term is added so that $\delta_p$ vanishes for $p\to\infty$, corresponding to the fact that the effects of the potential disappear at high energies. In fact only $l\ge 2$ gives $p\ge0$, while for $\tilde p$ we have $l\ge0$. Taking into account the two bound states, we have the same number of eigenstates with and without the kink. The determinant ratio in (\ref{eq:amplitude2}) at large $\tau_0$ evaluates to:
\begin{equation}
\label{eq:detratio}
\frac{\det\nolimits'{\left( -\square + V'' \right)}}{m^{-2} \det{\left(-\square + m^2\right)}} = \prod_{n=-\infty}^{\infty} \left\{\left(\frac{k_n^2}{\omega_n^2}\right)^{1 - \delta_{n 0}} \frac{k_n^2 + \frac{3}{4} m^2}{\omega_n^2} \prod_{l=2}^{\infty} \frac{p_l^2+\omega_n^2}{\tilde{p}_l^2+\omega_n^2}\right\}\,.
\end{equation}
We took into account that for $n=0$ the first bound state of the kink theory is the zero mode which has been already factored out.

Performing the product over $n$ by means of the identity:
\begin{equation}
\frac{\sinh z}{z} = \prod_{n=1}^\infty \left(1+\frac{z^2}{\pi^2 n^2} \right) \,,
\end{equation}
we can write the result in the form
\begin{gather}
\frac{\det\nolimits'{\left( -\square + V'' \right)}}{m^{-2} \det{\left(-\square + m^2\right)}} = (m L)^2  e^{m L \left(\frac{\sqrt{3}}{2}-2\right) +L \Sigma+2b},\\
\Sigma =\sum_{l=2}^{\infty} (p_l^2+m^2)^{1/2}-(\tilde{p}_l^2+m^2)^{1/2}\,,\\
b =\log\frac{1-e^{-\frac{\sqrt{3}}{2} mL}}{(1- e^{-mL})^2}+\sum_{l=2}^{\infty} \log(1-e^{-(p_l^2+m^2)^{1/2}L})-\log(1-e^{-(\tilde p_l^2+m^2)^{1/2}L})\,.
\end{gather}
For $\tau_0 m\gg 1$ we can approximate the sums by integrals:
\beq
\Sigma= \int_0^\infty \frac{dp}{\pi}\, \delta_p \frac d{dp} (p^2+m^2)^{1/2}=m(2-3/\pi-1/\sqrt{3})+\rm{log.div.}\,,
\eeq
where the logarithmic UV divergence is canceled in the final answer by the counterterm in \reef{eq:action}.

\begin{figure}[h!]
\begin{center}
\includegraphics[scale=0.6]{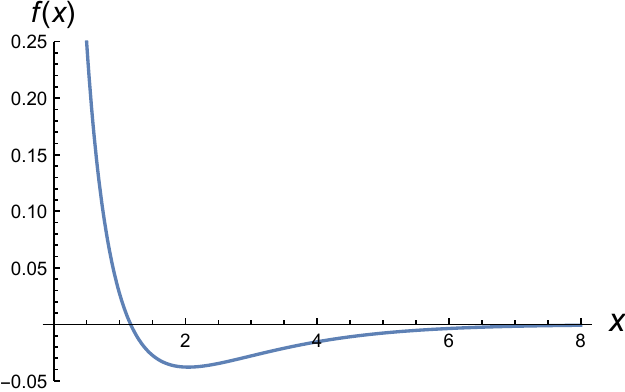}
\end{center}
\caption{The function $f(x)$ defined in Eq.~\reef{eq:f(x)}.}
\label{fig:f}
\end{figure}

Analogously 
\begin{align}
b&=\log\frac{1-e^{-\frac{\sqrt{3}}{2} mL}}{(1- e^{-mL})^2}+\int_0^\infty \frac{dp}{\pi}\, \delta_p \frac d{dp} \log(1-e^{-(p^2+m^2)^{1/2}L})=f(mL)\,,\\
f(x)&=\log(1-e^{-\frac{\sqrt{3}}{2} x})-\frac{2}{\pi}\int_0^\infty dq\, \biggl(\frac{1}{1+q^2}+\frac{2}{1+4q^2}\biggr) \log(1-e^{-(q^2+1)^{1/2} x})\,.
\label{eq:f(x)}
\end{align}
The function $f(x)$ tends to zero exponentially fast for $x\gg 1$, whereas for intermediate $x$ it has to be computed numerically, see figure \ref{fig:f}\,.

Gathering everything, the coefficient $\kappa$ in \reef{eq:amplitude3} is given by (cf.~\cite{Osborn:1978qg}, (3.27))
\begin{equation}
\kappa = \sqrt{\frac{m^3}{24 \pi g L}} e^{-L M_{\rm kink}-f(mL)} \,, \qquad M_{\rm kink} = \frac{m^3}{12 g} + m \left( \frac{1}{4 \sqrt{3}} - \frac{3}{2 \pi}\right)\,.
\end{equation}
Not surprisingly, the leading exponential dependence of this result is governed by the kink mass $M_{\rm kink}$ in the one-loop approximation, first computed in \cite{Dashen:1974cj}.

The one-instanton approximation for $\calA_-$ will break down for $\tau_0$ so large that $\kappa\tau_0=O(1)$. In this extreme $\tau_0 \to \infty$ limit, both amplitudes $\calA_\pm$ receive contributions from multi-instanton configurations in the path integral, which are approximate solutions of the equation of motion. We can use the instanton-gas approximations, where the centers of the instantons are far apart, and resum all these contributions, to give:
\beq
\label{eq:amplitudesgas1}
\calA_+ = \calA_0\cosh \kappa\tau_0,\qquad
\calA_- = \calA_0 \sinh \kappa\tau_0\,.
\eeq
We did not consider the purely perturbative corrections to these amplitudes, as they are the same for the quasi-degenerate states and therefore do not interest us.

Taking the $\tau_0 \to \infty$ limit in (\ref{eq:amplitudesgas1}), one can infer the presence of two exchanged states split in energy by
$\Delta \calE = 2 \kappa$, which is our final result.

\footnotesize

\bibliography{phi4-Biblio}
\bibliographystyle{utphys}

\end{document}